# Concept of an achromatic stellar coronagraph and its application for detecting extrasolar planets

**Bhavesh Jaiswal\* a, b**

a Space Astronomy Group, U R Rao Satellite Centre, ISITE Campus, Karthiknagara, 560037, Bengaluru, India.
b Indian Institute of Science, 560012, Bengaluru, India.

\*email: bhavesh@ursc.gov.in

**Imaging the planets that orbit around other stars requires blocking the host star which is usually 8-10 orders of magnitude brighter than the planets. This is achieved with the help of a stellar coronagraph. In the current work, a concept of a new type of stellar coronagraph is introduced where the star light is blocked by a linear polarizer in the collimated beam. It is based on differential rotation between the linear polarization state of planet light and that of star light. This is achieved with the help of a set of thick birefringent crystals in the collimated beam of a telescope where the planet light is made to travel extra optical path length compared to star light. By adjusting the orientation and thickness of the crystal, the optical path length can be made to cause a phase difference of π, just enough to rotate the initial plane of polarization by 90° for planet-light without affecting the star light. Theoretical calculations involving the phase difference due to birefringent crystals are presented here along with the basic configuration and design. It is shown that the design blocks the star light identically at all wavelengths. Application of this concept for detecting Earth-like extrasolar planet is discussed using a one-meter class telescope.**

Keywords: Polarimetry, Stellar Coronagraph, Exoplanet

1. Introduction

More than 4000 extrasolar planets have been discovered till date and most of these detections have been made using indirect techniques like the Transit and Radial Velocity method. These techniques have higher detection probability for the planets whose orbital planes are oriented towards the observer. A direct method of detecting extrasolar planets will involve imaging the planetary system around the host star, which will reveal the orbiting planets irrespective of their orbital orientation. Apart from discovering the planets, imaging also allows us to measure the spectroscopic and polarimetric signatures of reflected and emitted light from these planets which will facilitate the study of the atmospheric and surface signatures. Imaging an extrasolar planetary system using a telescope is challenging, primarily due to the brightness of the host star which overwhelms the light from any planet orbiting at small angular separations.

In order to detect an Earth-like planet orbiting a Sun-like star, hypothetically located at a distance of 10 parsecs, a stellar coronagraph would be required which can supress the star light by ~ $10^{10}$ times at a small angle of 0.1″ in the visible-NIR wavelength range (Ref. 1). To study the spectroscopic signatures of the Earth-like planet requires achieving this suppression (or nulling, blocking) over a broad wavelength band, typically visible to NIR. Several types of coronagraph designs have been invented for this purpose. Some designs work based on the principles of interferometry whereas others use the principles of Lyot coronagraph. The Interferometric Coronagraphs like Achromatic Interfero Coronagraph (Ref. 2) are based on interfering two beams of star light with a π phase difference to create a destructive interference. Recent results from ground-based nulling interferometry can be found in Ref. 3. The nulling interferometry using birefringent crystals (Savart plates) is attempted by Ref. 4. The pupil apodizing coronagraphs like Conventional Pupil Apodization (Ref 5) and Phase Induced Amplitude Apodization (Ref 6) can modify the phase or amplitude of the light in the pupil plane itself in such a way that it causes a null region at the focal plane. Lyot amplitude mask Coronagraphs like Band Limited coronagraphs (Ref 7) removes a major part of star light at the focal plane using a physical stop. Lyot phase mask coronagraphs like Four Quadrant Phase Mask (Ref 8) and Optical Vortex Coronagraphs (Ref 9) act on the phase of the star light at the focal plane to remove it. There is another category of coronagraphs, called external occulters, which creates a region of stellar shadow in space using a large petal shaped occulting structure. A telescope can be placed inside this shadow region to detect the planets (Ref. 10). There are few more examples of the coronagraphs in each category and a more detailed discussion on their working principle and performance can be found in Ref. 11 and Ref. 12.

The size of the points spread function (PSF) of the telescope image at the focal plane is proportional to $\lambda/D$ (ratio of wavelength to telescope diameter) factor. The chromatic performance of coronagraphs is largely dependent on $\lambda/D$ factor of the telescope. Some of these concepts are highly dependent on lambda/D (especially the ones operating at the focal plane) and hence are chromatic whereas others have a weaker dependence and have a broad wavelength performance. Fig. 7 of Ref. 12 compares the broadband performance of various kinds of stellar coronagraphs.

Here, I introduce a concept which is referred to as *Angle-Sensitive Achromatic and Polarizing Coronagraph* or **ASAP coronagraph**, which is a stellar coronagraph that blocks the on-axis host star identically at all wavelengths and is suitable for detecting planets at small angles of separation from the host star. The ASAP coronagraph has the advantage that it does not demonstrate any $\lambda/D$ dependence (and hence achromatic) since the design allows blocking of the host-star light at the pupil plane itself, instead at the focal plane where the PSF has a strong $\lambda/D$ dependence. The other advantage is that the design can be used with a relatively small aperture telescope. The ASAP coronagraph polarizes the incoming star and planet light at the exit of the telescope secondary mirror and makes use of the angular separation between the host-star and its orbiting planet, to rotate the plane of polarization of the planet's light (without affecting the star light) with the help of birefringent crystals. The star light is then blocked by a linear polarizer while the planetary light is allowed to be transmitted by the same polarizer.

In this work, the ASAP coronagraph concept is introduced with the assumption of ray-propagation. In the following sections, section 2 describes the phase difference due to birefringent crystals and section 3 introduces the ASAP concept, along with its applications for detecting Earth-like planets. In section 4, some of the important design considerations are discussed with a conclusion in section 4.

## 2. Birefringent crystals and phase difference

Birefringent crystals (like Calcite, Quartz etc.) have two different refractive indices (ordinary and extra-ordinary refractive indices: $n_o$ and $n_e$ respectively) in two orthogonal directions called ordinary and extra-ordinary axes of the crystal (Ref 13). The electric field vector of the light ray entering the birefringent crystal experiences a refractive index depending upon its orientation with respect to the two axes. In a simple geometry of transmission, this phenomenon can cause a phase difference (or a phase lag) between two orthogonal electric field directions. The larger the optical path length travelled by the rays, the larger is the phase difference accumulated. Since the refractive indices are different for the ordinary and extra-ordinary rays, the refractive bending of the rays is also different. This causes a physical separation among the two rays in the direction of propagation.

For an arbitrary transmission geometry in a plane-parallel birefringent crystal of thickness $L$ as shown in Fig. 1 on left, the phase difference between extra-ordinary and ordinary ray is given by $\Delta\Phi = (\Phi_o - \Phi_e)$ where incidence direction is marked by angle $\alpha$ and optics axis is marked by two angles $\delta$ and $\Theta$. This crystal is considered to be immersed in a medium with refractive index $n$. From Eq. 12 of Ref. 14, $\Delta\Phi$, at the exit of the crystal, can be found to be

$$\Delta\Phi = \Phi_o - \Phi_e = \frac{2\pi L}{\lambda} \times \left[ (n_o^2 - n^2 sin^2\alpha)^{0.5} + \frac{n(n_o^2 - n_e^2)sin\theta cos\theta cos\delta sin\alpha}{n_e^2 sin^2\theta + n_o^2 cos^2\theta} + \right.$$

$$\left. \frac{-n_o[n_e^2(n_e^2 sin^2\theta + n_o^2 cos^2\theta) - \{n_e^2 - (n_e^2 - n_o^2)cos^2\theta sin^2\delta\}n^2 sin^2\alpha]^{0.5}}{n_e^2 sin^2\theta + n_o^2 cos^2\theta} \right]. \quad (1)$$

The phase difference gathered by a 50 mm thick calcite crystal for a small range of incident angles is seen on the right in Fig. 1. The wavelength considered here is 500 nm and angle Θ is kept as 0°. This configuration is same as that of the crystal C1 used in section 3. The phase difference is seen to be varying with angle α and not with δ. The variation with δ is very slow (about 100 times slower) compared to α mainly because angle variation about α=45° causes larger variation in the optical path length than about δ =90°.

As discussed in the next section, the incident ray to the crystal is linearly polarized in such a way that its electric field vector can be split equally into two orthogonal directions (those directions being the *o* and *e* axes of the crystal). So, the two rays having an orthogonal polarization originate from the same incident ray. In such condition, the different amount of phase gathered by the two rays inside the crystal can cause the outcoming ray to be in a state of polarization different than the incident ray. This is achieved by keeping the direction of polarization of the incident ray to be at 45° from the *o* and *e* directions. If this is the case, then the outcoming ray can be linearly, elliptically or circularly polarized depending upon the phase difference.

3. **Concept design of the ASAP coronagraph and challenges involved**

Consider a telescope with a primary and a secondary mirror of focal lengths *f1* and *f2*. The beam at the exit of secondary mirror is a collimated beam (with a flat wavefront). The angular separation between planet and star light, after the secondary mirror, is $\epsilon$, which is the magnification (=*f1/f2*) of the telescope multiplied with the initial separation angle. The choice of *f1* and *f2* is mainly responsible for the reduction in the beam diameter by the factor *f1/f2* so that the initial star-planet angle can be increased. Even if this combination of focal lengths is not available one can use additional optics to increase or reduce the beam diameter for the same effect. When entering the ASAP, the star and planet-light are considered to be unpolarized. Consider a coordinate reference frame with *z* axis in the direction of light propagation. At the exit of the secondary mirror a linear polarizer is used to polarize both the star and planet light. The polarizer axis is at 45° with respect to *x* and *y* axes. If we denote star by subscript '0' and planet by '1' then, after the polarizer $(\Delta\phi)_0 = 0$ and $(\Delta\phi)_1 = 0$ (here $\Delta\phi$ is the phase difference between two orthogonal linear polarization states parallel to *x* and *y* directions given by $\phi_X - \phi_Y$) as the polarizer itself does not add any relative phase. After the polarizer, the two beams (star and planet light) travel in a plane parallel birefringent crystal plate with a very small angle $\epsilon$ between them. The star and planet beams gather a phase difference of $(\Delta\phi)_0$ and $(\Delta\phi)_1$ respectively, due to their different optical path lengths inside the crystal. The objective here is to make a configuration where the two beams travel their respective distances such that we achieve the condition of

$(\Delta\phi)_0 = 2m\pi$     &     $(\Delta\phi)_1 = (2m+1)\pi$.      (2)

Here *m* is an integer and is also called 'order'. This condition ensures that if incident beams (star and planet) are initially polarized in the same plane then the planes of polarization of the outcoming beams (star and planet) are orthogonal to each other. One of beam can be blocked (star) and other can be transmitted (planet) by using a linear polarizer. Now, there are three major challenges at this stage which prevent us from achieving this condition:

(i) to achieve this condition, it would require physically impractical lengths of the crystal for very small $\epsilon$.
(ii) even if condition (i) is achieved phase difference will be varying very fast with wavelength due to presence of $L/\lambda$ term in Eq 1. Since $L>>\lambda$, the performance will be changing very rapidly with wavelength for even few millimeters of length $L$.
(iii) as shown in Fig. 1, the extra-ordinary and ordinary beams start to separate after passing through crystal and this would imply that incident beam (for star as well as the planet) will decompose into two beams travelling at a small angle to each other and may not overlap anymore. This means that in a collimated beam, only the region where two beams overlap will be useful.

First challenge is partly solved by the magnification of the telescope (which can increase $\Theta$ multi folds) and partly by tilting the crystals (large values of angle $\alpha$ in Fig. 1) which allows to achieve this condition within the reasonable lengths of the crystal. Tilted crystals can cause minor dispersions which can be compensated at a later stage using a glass wedge (and is not discussed here). Second challenge is solved by passing the light through two identical crystals whose axes have mutually perpendicular orientation (like a Babinet-Soleil compensator). This configuration will make $m$ in Eq. 2 equal to 0 and hence makes is perfectly achromatic for star light, though there is still a very slow variation on wavelength of off-axis planet light. The phase gathered by star light in one crystal is compensated by the other crystal and hence star light gathers no extra phase difference in its $x$ and $y$ components. This means that the state of polarization of star light, which is perfectly on-axis light, remains same as initial (for all wavelengths) due to equal optical path length travelled inside two crystals. However, the direction of polarization of the off-axis light starts to gather extra phase due to unequal travel distance in the two crystals. This extra phase gathered by the off-axis ray allows for the change in the direction of polarization. This causes the planet light, having an extra phase of π, to be rotated by 90° with respect to initial direction of polarization. See the inset of Fig. 3 for variation of state of polarization from linear to elliptical to linear again as the phase difference increases for the off-axis light. The third challenge is solved by using another set of identical crystals whose axes are oriented such that they correct for the beam displacement. One can do away with this challenge if the beam diameter entering the coronagraph is large and hence causing a good overlap in the extra-ordinary and ordinary beams at the crystal exit. However, by doing this there will be some loss of throughput since only the overlapping region of the beam can be useful (similar to Ref. 4).

A complete configuration of this coronagraph setup is shown in the following Figure 2 (telescope primary and secondary are not shown). The incidence angle for star light beam is $\alpha=45°$ and for planet light beam $\alpha=45°+\epsilon$. The coordinate reference frame and orientation of optical axes and polarizer axes are also marked for reference. Crystal (C1) is rotated by 45° about X axis and the second crystal (C2) is rotated by 45° about Y axis. Identical crystals C1 and C2 are used in order to keep the system achromatic, as discussed above in challenge (ii). After C1 and C2, two more crystals C3 and C4 (identical to each other) are used in order to achieve a good overlap between the beams, as discussed above in challenge (iii). Both beams are polarized initially (at plane A) and at the output (at plane B) the polarization of planet beam is seen to be rotated by 90°. After plane B, a linear polarizer P2 is used to block the star light which leaves only planet light at plane C. After blocking the star light in the collimated beam an image can be formed using a lens or a mirror after plane C.

Following section discusses in detail on how this concept can be used in a real telescope system for detecting extrasolar planets at close angles.

### 3.1. Application of ASAP coronagraph for detecting extrasolar planets

For a Sun and Earth like system situated at 10 parsecs the angular separation between them is 0.1″. In order to achieve condition in Eq. 2 for such a small angle it would require impractically large thicknesses of crystals (about a meter). Instead, this small angle can be increased to ~100 times by using the magnification of the telescope. A combination of primary mirror and telescope optics can be chosen such that the beam diameter is reduced by 100 times (that is, assuming a 1.5 m telescope primary mirror and a beam incoming to the ASAP with a diameter of 15 mm, the star-planet angle is increased to 100 times) and increase the initial star-planet angle is increased to 10″ (= ϵ). For this angle, condition 2 can be achieved (with $m=0$) by using two identical 50 mm (=$L$) thick crystal of Calcite (as in Fig 2) with incidence angle of 45°. Using Eq. 1 for this geometry, one can find $(Δϕ)_0 = 0$ (with $α = 45°$) and $(Δϕ)_1 = 180°$ (with $α = 45° + 10″$) after the ray passes through C1 and C2. To get the total phase after C1 and C2, one has to add the individual phases gathered by light in C1 and C2. In terms of Jones calculus, the complex amplitude of electric field in the *X* and *Y* directions can be written as:

$$\begin{bmatrix} E_x \\ E_y \end{bmatrix} = \begin{bmatrix} E_0 e^{i\varphi_x} \\ E_0 e^{i\varphi_y} \end{bmatrix}.$$

It is to be noted that physical Electric field is the real part of this vector.

The Jones vector of the polarized light, after polarizer P1, can be written as $\frac{1}{\sqrt{2}}\begin{bmatrix}1\\1\end{bmatrix}$ where P1 is oriented at 45° with respect to positive X-axis. Writing the Jones matrices of C1 and C2 in terms of X and Y directions, we get:

$$J_{C1} = \begin{bmatrix} e^{i\varphi_{XC1}} & 0 \\ 0 & e^{i\varphi_{YC1}} \end{bmatrix}$$

and $J_{C2} = \begin{bmatrix} e^{i\varphi_{XC2}} & 0 \\ 0 & e^{i\varphi_{YC2}} \end{bmatrix}$. Since the extra-ordinary axis of crystal C1 and C2 is in the *X* and *Y* directions respectively, the Jones matrix of C1 and C2 can also be written as follows:

$$J_{C1} = \begin{bmatrix} e^{i\varphi_e} & 0 \\ 0 & e^{i\varphi_o} \end{bmatrix}$$

and $J_{C2} = \begin{bmatrix} e^{i\varphi_o} & 0 \\ 0 & e^{i\varphi_e} \end{bmatrix}$.

In order to relate it with eq. 1, we can see that $\phi_Y - \phi_X = ΔΦ_{C1}$ for C1 and $\phi_X - \phi_Y = ΔΦ_{C2}$ for C2. A combined Jones matrix of C1 and C2, obtained after the product of the two individual matrices is given by:

$J_{C1} \times J_{C2} = \begin{bmatrix} e^{i\varphi_{XC1} + i\varphi_{XC2}} & 0 \\ 0 & e^{i\varphi_{YC1} + i\varphi_{YC2}} \end{bmatrix} = e^{i\varphi_{XC1} + i\varphi_{XC2}} \begin{bmatrix} 1 & 0 \\ 0 & e^{iΔ\varphi_{C1} - iΔ\varphi_{C2}} \end{bmatrix}$. Since we are interested in the relative phase between X and Y direction of Electric field, we can ignore the term outside the matrix which is common to both X and Y directions. Combining the Jones matrices of C3 and C4 also in this term yields the combined matrix of all 4 crystals together: $J_{C1} \times J_{C2} \times J_{C3} \times J_{C4} = \begin{bmatrix} 1 & 0 \\ 0 & e^{iΔ\varphi_{C1} - iΔ\varphi_{C2} + iΔ\varphi_{C3} - iΔ\varphi_{C4}} \end{bmatrix}$. The Jones matrix of the polarizer P2 (oriented at -45°) is given by $\frac{1}{2}\begin{bmatrix} 1 & -1 \\ -1 & 1 \end{bmatrix}$. Finally, the light coming out after P2 is given by

$$\begin{bmatrix} E_x \\ E_y \end{bmatrix} = \tfrac{1}{2} \begin{bmatrix} 1 & -1 \\ -1 & 1 \end{bmatrix} \times \begin{bmatrix} 1 & 0 \\ 0 & e^{i\Delta\varphi_{C1} - i\Delta\varphi_{C2} + i\Delta\varphi_{C3} - i\Delta\varphi_{C4}} \end{bmatrix} \times \tfrac{1}{\sqrt{2}} \begin{bmatrix} 1 \\ 1 \end{bmatrix}.$$

$$\begin{bmatrix} E_x \\ E_y \end{bmatrix} = \tfrac{1}{2\sqrt{2}} \begin{bmatrix} 1 - e^{i\Delta\varphi_{C1} - i\Delta\varphi_{C2} + i\Delta\varphi_{C3} - i\Delta\varphi_{C4}} \\ -1 + e^{i\Delta\varphi_{C1} - i\Delta\varphi_{C2} + i\Delta\varphi_{C3} - i\Delta\varphi_{C4}} \end{bmatrix} \qquad (3)$$

Total intensity after the polarizer P2 becomes $|Ex|^2+|Ey|^2$. For the starlight the combined phases are 0, and then we find that $|Ex|^2+|Ey|^2 = 0$. For the planet light $|Ex|^2+|Ey|^2 = 1$ as the combined phases are equal to π.

For a 50 mm thick crystal, the *e* and *o* beams get separated by about 3 mm after C1 and C2 as shown in Fig. 2. To correct for this separation, two crystals C3 and C4 of similar thickness are used as shown in Fig. 2 with optic axes oriented at angle of ~ 18° (in X-Z and Y-Z plane respectively) with respect to the direction of light propagation. It is to be noted that the sole purpose of C3 and C4 is to shift the beam spots, but they also add a phase $\Delta\phi$ to the rays. The angle α is equal to 10˚ for C3 and C4. The beam shift caused by crystal C1 is corrected by C4 and that caused by crystal C2 is corrected by C3. It is to be noted that the direction of optic axis is perpendicular to the plane of refraction in C1 whereas it is parallel to the plane of refraction in C4. The choice of *α*, *L* and optic axes direction are inter-dependent and should be chosen accordingly. The optic axis direction is chosen here to achieve a perfect beam overlap within the crystal of given thickness (50 mm). It is advisable to use tilted crystals (*α>0*) to avoid the multiply reflected beam entering the beam path and hence we choose *α=10°*. It is possible to choose other combinations of *α*, *L* and optic axis direction for C3 and C4 as long as both the crystals are identical and symmetrically placed. One can use Eq. 1 to calculate phase difference and proceed as discussed above to calculate the phase difference in Ex and Ey coming out after P2. Table 1 gives the values of various parameters used here for the calculations.

The on-sky pattern of complete phase differences of two crystal pairs, that is $\Delta\Phi_{C1}$ -$\Delta\Phi_{C2}$ and $\Delta\Phi_{C3}$ -$\Delta\Phi_{C4}$ (of Eq. 3) are shown, in radians, for two wavelengths in Fig. 3 on left and middle panels respectively and the intensity pattern after P2 is shown in a similar manner on right. The angle scales consider a planet star-system located at 10 parsecs. The star is shown the centre and the orbit of a planet at 1 au separation is marked. As can be seen, the phase difference at the centre (at the star) is always 0 and hence the star light is having the same state of polarization, irrespective of wavelength. As the angle moves away from the centre the phase difference ($\Delta\phi$) starts to get finite values which causes the state of polarization to change from linear to elliptical to circular and then change the direction and again become elliptical and then linear. This transition of state of polarization is represented in the inset of Fig. 3. The magnitude of phase difference accumulated due to the two crystal pairs can also be seen. The phase accumulated by the two crystal pairs contribute to the total phase difference ($=\Delta\Phi_{C1}$ -$\Delta\Phi_{C2}$ + $\Delta\Phi_{C3}$ -$\Delta\Phi_{C4}$).

Considering the phase difference diagram in Fig. 1 (right) it can be seen that the phase difference between extra-ordinary and ordinary beam, $\Delta\Phi$ for C1 will have a linear variation only in the Y direction and for C2 will have a linear variation only in the X direction due to the large tilt of the crystals in Y and X directions respectively. Now a combination of C1 and C2 together, as used in this concept, gives a phase difference in the direction diagonal to X and Y directions (as in Fig. 3, left panel) and that is why the intensity fringes are seen to be inclined at 45˚ with respect to horizontal direction. For large δ (typically ~100 times than considered here), the phase difference as well as the intensity fringes will start to curve owing to the variations in the azimuthal angle (δ).

The on-sky pattern is shown as bright and dark fringes and to block the star light it is required to keep the star on a dark fringe. Also, it may be required to either rotate the spacecraft or the coronagraph (setup in Fig. 2) inside the spacecraft about the pointing axis to search for the planets in all locations around the star so that the bright fringe intersects the planetary orbit at all locations (a strategy like the spinning interferometer concept of (Ref. 15)).

### 3.2. Contrast achieved for a finite star-size and a realistic polarizer

As discussed earlier, the contrast required to image Earth-like planets around Sun-like stars is about ~$10^{-10}$. The star's light is required to be suppressed by this factor. In order to achieve this contrast with ASAP we need to consider two additional important factors which we have ignored in the discussion so far: (i) the extinction ratio of linear polarizers and (ii) finite size of the stellar disc. So far, we have considered the polarizer as an ideal linear polarizer and the star as a point source. In a realistic scenario, the best available polarizers do not have extinction ratio larger than $10^6$:1 and the size of the stellar disc is close to 0.5 milli-arcsec for a Sun like star located at 10 parsecs. To consider a realistic polarizer we take help of Mueller calculus as it can account for partially polarized light as opposed to Jones calculus. The calculations are similar to Jones calculus as mentioned above and are discussed in the appendix. One can refer Ref. 13 for details of the Mueller matrix representation. The stellar suppression for a realistic polarizer in ASAP system is shown in Fig. 4. As can be seen, using an ASAP coronagraph configuration with a realistic polarizer of extinction ratio of $10^6$:1, one cannot achieve a contrast better than $10^6$. However, the design of this coronagraph offers the advantage of improving the contrast by using more coronagraphs in series (like Ref. 16). This can be helpful to remove any star light leakage due non-zero phase of star light which can be caused by a finite angular size of the star. Using more coronagraphs in series will also help in blocking the star light beyond $10^{-6}$ level as it will allow the star light to go through a greater number of sets of crossed polarizers (in the single stage as in Fig. 2, the star light goes through only one set of crossed polarizers, that is, P1 and P2 and hence the extinction better than $10^{-6}$ is not possible). The simulations considering several ASAP stages in series (also discussed in appendix) show that the contrast of $10^{-10}$ can be achieved by using 3 or more number of stages. The optical setup depicted in Fig. 2 is to be considered as a single stage coronagraph. The following Fig. 4 shows the intensity contrast of the fringes observed in Fig. 3 for 1, 2, 3 and 4 stages combined in series. As can be seen in the figure, the intensity drop is very sharp with angle and the star (angular diameter shown with vertical line at the centre) needs to fall just at the centre of this cusp. The size of the stellar-disc considered here is also multiplied with telescope magnification (considered to be 100). Simulations are shown for two wavelengths, 500 and 1000 nm. As is also seen in Fig. 3, the width of the fringe pattern also increases with wavelengths.

Fig. 5 shows the second stage, as a continuation of single stage of Fig. 2, which has to be used after the first stage. The residual star light after P2 is shown as dashed line whose plane of polarization is same as that of planet-light. Although the star and planet light before P2 are seen to be in mutually perpendicular plane of polarization in Fig. 2, the residual star light in Fig. 5 is shown in the same plane of polarization as planet light. This happens as due to the finite size of the star, the light of the stellar limb incident at P2 is not linearly polarized but slightly elliptically polarized and the component of that ellipse, with its plane of polarization in the same direction as planet light, acts as a residual star light. The light coming from the central part of the star gets completely removed after P2. Even though after P2 the residual star light has same plane of polarization as that of planet light, both the rays still maintain the same angular difference with each other which is why use of a second stage (similar to first stage) has the same effect in again reducing the stellar light by the same amount as first stage. After using the second stage this residual star light gets rotated after going through crystals C1-C4 and then another polarizer P1 reduces the star light without compromising on the planet-light. Each consecutive stage is supposed to reduce the star light by the same factor, however as we see in Fig. 4 that for a realistic polarizer (having a finite extinction ratio) single and double stages have same stellar suppression of $10^{-6}$ and only after adding third stage the suppression reduces to $10^{-12}$. This happens as for non-ideal polarizer we need a 'pair' of crossed polarizers to consecutively reduce the star light. Each stage has only one additional polarizer and hence it would require 3 stages (totally 2 'pairs' of crossed polarizers) to enhance the suppression from $10^{-6}$ to $10^{-12}$. One can convince themselves of this step with the help of simple matrix multiplication of Mueller matrices of polarizers as explained in the Appendix below.

### 4. Discussion towards a practical implementation of ASAP coronagraph

There are several considerations which need to be brought out for a practical implementation and understanding of ASAP concept of coronagraph. Some of these considerations are advantageous and some can be challenging for performance and need to be understood in detail.

**Multiple reflections within the crystal:** The plane parallel crystals used here can have multiple reflections of light within the parallel walls of the crystal. Although most of the light is transmitted out of crystal but some small part of light can always undergo multiple reflections and can be detrimental to the design due to un-accounted additional phase which it gathers via multiple reflections. More reflections will add more phase in the ray and ultimately have effect on the contrast achieved with this coronagraph. Thankfully, the use of thick ($L$=50 mm) and tilted crystals ($\alpha$=45°) can take care of this problem if the beam diameter is not too large. After the first internal reflection itself the internally reflected beam can be made to go out of path of the main beam. The multiple reflected rays will come out from the crystals at a lateral separation from the transmitted rays which can then be blocked by a circular aperture.

**Length $L$ of crystal and angle $\alpha$:** The length or thickness $L$ of the crystal and incidence angle $\alpha$ have similar effect on the performance of the coronagraph. Increasing $L$ or $\alpha$ increases the close sensing capabilities of this coronagraph. One can be used to trade-off the other in the design. A careful selection of length and incidence angle can bring the limits of planet detection, closer to or farther away from the host star. The ASAP concept is demonstrated above in section 3.2 for a choice of 'L' and '$\alpha$' which permits the closest separation of planet at 0.1″, however by changing these two parameters one can alter this limit. In the following Fig. 6 are shown the intensity fringes for nine different combinations of 'L' and '$\alpha$'. Larger values bring the detection zone closer to the star and vice versa.

**Tolerances on length $L$ of crystal and angle $\alpha$:** The ASAP technique introduced here uses two identical birefringent crystals (C1 and C2) to cause a phase difference with angle. If due to manufacturing processes, there is a small mismatch in the thickness of two crystals it will cause a constant shift in the phases gathered by star and planet light (although they both will still differ by π) and hence the output of C2 may not be linearly polarized light. To correct for this extra constant phase, one can consider using a Babinet-Soleil compensator, which can produce a 'tunable' extra optical path length even for single order, it would be possible to add extra phase in one of the directions of polarization. It can be inserted in the optical chain after C2 and can be tuned to compensate for the extra phase offset.

To mimic the mismatch in length and angles of first pair of identical crystals, the phase difference $\Delta\Phi_{C1}$ -$\Delta\Phi_{C2}$ is calculated by deliberately adding an extra length to 'L' and an extra angle to '$\alpha$' of crystal C1.

This phase difference is shown in Fig. 7 (left). Considering the size of stellar disc, the desirable phase difference which can be tolerated is about ~ 0±0.015 radians. This region is marked by a contour of two parallel lines in the plot. As can be seen from this plot, the ASAP-coronagraph is very sensitive to mismatch in the incident angles and thickness of two crystals. A desirable accuracy for C1 and C2 in the incident angle and thickness can be seen from the figure to be about ~0.1″ and ~10 nm respectively. An increase or decrease in the length or angle can cause a proportional change in the phase difference. Also, one can see that an offset in one (angle or length) can be compensated by the other. It will be required to either have a precise control on the angle of one of the crystals or the length of one of the crystals by inserting a Babinet-Soleil Compensator (composed of a set of birefringent crystals as shown on Fig. 7, right). This compensator can be tuned (or calibrated) to cause an extra path length for ordinary or extra-ordinary polarization by 'sliding' one of the crystals in a controlled manner and increasing the optical path. The similar tolerance will be valid for crystals C3 and C4 too.

**Telescope Pointing Errors:** Although this design is insensitive to lateral shift of the beams in X or Y directions but is very sensitive to the angular shifts or errors in the telescope pointing. Since it works in the collimated beam which is also magnified, any error $\varDelta\alpha$ in telescope pointing will cause $M*\varDelta\alpha$ change in the angle inside the coronagraph, where $M$ is the magnification of the telescope. The sensitivity to pointing is maximum in the direction perpendicular to the direction of fringes (shown in Fig. 3) and zero in the direction parallel to the fringes. Adding more stages can be beneficial as it will broaden the 'null' or dark region of the fringes so that the small pointing errors can be accommodated in the broad 'null' region.

**Chromaticity:** The ASAP design ensures m=0 in Eq. 2 which allows the outcoming polarization states of all the wavelengths to be the same as incident and hence all the wavelengths in the on-axis directions are blocked. Even though there is a variation of refractive index with wavelength, the m=0 condition allows for total phase to be zero for broad wavelength range for the on-axis light. Although the star light is blocked at all wavelength identically, the planet light (or the bright fringes in Fig. 3), having m=1, 2 and so on for consecutive fringes, shows slow variations with wavelength. The fringe spacing (or angular resolution) increases linearly with wavelength. For a crystal length of 50 mm and $\alpha = 45°$, it gives a broad wavelength performance (for all $\lambda< 1000$ nm) at separation angle of 0.1″. The chromaticity in the refractive index does not significantly affect the performance of the coronagraph but does cause a small chromatic beam shift of ~ 300 μm for C1 and C2 and ~ 150 μm for C3 and C4 over a broad wavelength range of 500-1000 nm. The chromatic beam shifts are caused here mainly due to the dispersion of light at the refraction boundary. Due to difference in the relative orientation of the optic axis in C1 and C4 (also in C2 and C3) crystals the dispersion is not the same in the two crystals and this leads to a difference in the chromatic beam shifts.

**Fresnel propagation of Beam:** The ray optics consideration of light propagation considers the light beam to be completely collimated (a flat wave front) before entering ASAP. However due to the diffraction effects, encountered due to the edges of primary mirror, the entering wave front may not be a perfect flat wave front. The phases of Electric field vectors on this propagating wave front will change with distance. This has to be calculated considering the Fresnel propagation of the wave front. The non-planarity of the propagating wave front is a serious problem for other coronagraphs too. However, there are remedies to overcome this problem like pupil-apodization where the edges of the pupil can be masked (as in Ref. 17) or using deformable mirrors (as in Ref. 18) and thus reducing the major sources of this non-planar wavefront. Since ASAP method works by using the polarized light and not just the intensity, a careful analysis has to be done first by considering the two polarizations separately and then for a combined scenario where the phases of the two polarizations are added.

**Choice of birefringent crystal and working wavelength range:** The phase gathered by a ray travelling inside a birefringent crystal is directly proportional to the birefringence ($\Delta n = n_e - n_o$) of the material. In this work we have considered an example of Calcite which has a birefringence of $\sim -0.17$. There are other common birefringent crystals like Quartz, but they cannot be used effectively for this concept if they have a very low birefringence ($\Delta n$ of Quartz is 0.009) as it will require impractically large lengths of such crystals. Rutile can be considered as another choice of material for this concept which has a large birefringence of $\sim 0.28$. Note that the sign of birefringence has no effect on the working of ASAP concept. Considering the transmission of crystals, Calcite can be a good choice for 400 – 2000 nm wavelength range and Rutile can be considered for 2000 – 4000 nm wavelength range.

Refractive index of any material has a slow variation with wavelength and can be obtained via Sellmeier equation. For example, in case of calcite, the birefringence is seen to vary from -0.177 to -0.163 for a wavelength of 500 and 1000 nm respectively. This variation (<10% in $\Delta n$) has a proportional effect on the fringe spacing of Fig. 3 and 6. It is to be noted that the variation in refractive index with respect to wavelength has its effect only for the off-axis planet light and not on the on-axis star light as discussed in section 3 (because m=0 for on axis light and m=1 for the first bright fringe) and hence making it achromatic.

**Throughput of the coronagraph:** ASAP coronagraph introduced here works by initially polarizing the planet and star light and then causes a relative phase difference of $\pi$ in the planet light and thereby rotating the initial direction of polarization of planet light. Since the light is linearly polarized (after P1) before it enters the birefringent optics, there is an absolute intensity reduction by 50% in ASAP concept. This can be detrimental for the performance of the coronagraph, especially for detecting smaller planets which are already photon starved. To overcome this problem, it is suggested to split the incident unpolarized light (before P1) into two orthogonally polarized beams using a polarizing beam splitter. If the light is split before P1 into two orthogonally polarized beams then it would be possible to have two parallel arms (from P1 to C) of the coronagraph setup (Fig. 2), each having a feed of the polarized input beam after the beam splitter. The two images can be recorded separately by two detector or can be combined and imaged onto same detector and hence without any compromise on the planet signal. Also, to be noted here is that combining several stages reduces only the signal of the star but has no effect on the planet signal. There can be other losses of the throughput which can arise due to reflection from the crystal surface or absorption through the crystal. These are unavoidable losses in any transmission optics but can be minimized with the help of anti-reflective coatings on the crystal and other optical surfaces.

5. **Conclusion and Future directions**

In this work, an innovative approach to high contrast stellar coronagraph design by the method of ASAP coronagraph is presented. This approach works by applying a differential rotation of state of polarization of incident light with respect to the angle and then selectively blocking the star light while filtering out the planet light. It works by passing the light through two birefringent crystals of equal lengths. ASAP coronagraph blocks the stellar light in the collimated beam of the telescope itself by carefully maintaining the optical path lengths inside the crystal and hence is closely related to nulling interferometry technique. A laboratory demonstration of ASAP coronagraph concept is planned which will help in establishing the methodology and in studying the sources of errors. Laboratory test will also help in finding the sources of unwanted extra phase which can arise due to:

- Difference in the temperature of two identical crystal leading to variations in refractive index. The temperature variation of refractive index of calcite crystal is studied by Ref. 19 and it is

- observed that the variation is about ~ 0.0001 per °C. It may be required to maintain the temperature of the two crystals very close to each other, else it may also cause an extra phase (similar to Fig. 7) which will then need to be corrected.
- Differences in the two 'identical' crystals C1 and C2 which may be caused due to variations (inhomogeneity) in optical quality of the crystals. The extra phase added due to these differences may need to be calibrated and then compensated.

Also, to be noted here is that the reflected light from the extrasolar planet can itself be polarised owing to Rayleigh and Mie scattering of light in the atmosphere of the planet. The degree of polarization can be as high as 10-20% in visible and NIR wavelength range (Ref. 20). The degree of polarization as well as orientation of polarization can be dependent on the phase angle of the planet. In such a scenario, it can improve or reduce the SNR (signal to noise ratio) depending upon the relative orientation of P1 with respect to star-planet system. An a priori knowledge of the planet location can help in re-orientation of the satellite in order to maximize the planet signal entering ASAP coronagraph.

The calculations presented in this work are limited to ray-optics calculations. To make the calculations more robust it will be required to consider the Fresnel propagation in the collimated beam and verify the performance. The wave front irregularities due to beam propagation, surface deformities and wavefront control etc. are some of the effects which need to be studied in future work and can affect the performance. The proposed method has similarities to interferometric methods (Ref. 2) that it removes the light in the pupil plane itself by using precise control of optical path length to achieve a phase matching condition. It also has similarities to polarization interferometry in terms of polarised fringes using birefringent crystals (Ref. 4). A comparative study with these existing concepts will be helpful in future to have a better understanding of this method.

The ASAP coronagraph, which can remove the bright light of the star in the pupil plane of the telescope, can work equally well for single mirror or segmented mirror telescopes and hence serves as a potential backend instrument for ground-based as well as space-based telescopes. It is needed to explore this concept further for future detection and characterization of extrasolar planets.

*Acknowledgement: I thank the two anonymous reviewers for giving a set of very useful comments and suggestions. I would like to thank Dr Sankar, SAG, for some useful discussions while developing this concept.*

**Appendix:**

In order to simulate a realistic system, we take help of the Mueller calculus which, unlike Jones calculus, is capable of handling partially polarized light. For a perfect linear polarizer, the component of electric field parallel to polarizer axis is completely transmitted whereas the component orthogonal to polarizer axis is completely blocked. Following Ref. 21 (Section 4.6.3), for a partial linear polarizer, the two orthogonal components $E_X$ and $E_Y$ are assumed to be affected by two positive constants $k_1$ and $k_2$ respectively. For a perfect polarizer oriented along X-axis, $k_1=1$ and $k_2=0$. The Mueller matrix of a partial polarizer is then given as,

$$P = \frac{A}{2}\begin{bmatrix} 1 & Bp_2 & Bq_2 & 0 \\ Bp_2 & p_2^2 + Cq_2^2 & (1-C)p_2q_2 & 0 \\ Bq_2 & (1-C)p_2q_2 & q_2^2 + Cp_2^2 & 0 \\ 0 & 0 & 0 & C \end{bmatrix}.$$

Here, $A = k_1^2 + k_2^2$, $B = (k_1^2 - k_2^2)/A$, $C = 2k_1k_2/A$, $p_2 = \cos 2\omega$ and $q_2 = \sin 2\omega$, where $\omega$ is the angle of orientation of the polarizer axis with the horizontal direction (in our case, $\omega = \pm 45°$).

The linear polarizers are often characterized by their extinction ratios, which is defined as the ration of the light passing through a pair of parallel polarizers to that of the light passing through a pair of crossed polarizers. For a perfect polarizer it is infinitely large but the extinction ratio of available crystal polarizers can be ~$10^6$:1. According to the definitions described above, this extinction ratio = $(k_1/k_2)^2$ =$10^6$. The square sign appears when converting the electric field amplitudes to Intensity. Now, assuming $k_1$=1, for a perfect transmission along the polarizer axis, we get, $k_2$=0.001, A=1.000001, B=0.999998 and C=0.001999. Thus, the Mueller matrix of this realistic linear polarizer oriented at ±45° is given as follows:

$$P(\pm 45°) = \frac{A}{2}\begin{bmatrix} 1 & 0 & \pm B & 0 \\ 0 & C & 0 & 0 \\ \pm B & 0 & 1 & 0 \\ 0 & 0 & 0 & C \end{bmatrix}.$$

The Mueller matrix of the birefringent crystals (see Ref. 13, 21) can be written as:

$$M = \begin{bmatrix} 1 & 0 & 0 & 0 \\ 0 & 1 & 0 & 0 \\ 0 & 0 & \cos\sigma & \sin\sigma \\ 0 & 0 & -\sin\sigma & \cos\sigma \end{bmatrix}.$$ Here, σ is the phase difference = $\Delta\varphi_C$, as described above. One can see after a matrix multiplication that the Mueller matrix of all the 4 crystals together is given by

$M = C1 \times C2 \times C3 \times C4 =$
$$\begin{bmatrix} 1 & 0 & 0 & 0 \\ 0 & 1 & 0 & 0 \\ 0 & 0 & \cos(\Delta\varphi_{C1} - \Delta\varphi_{C2} + \Delta\varphi_{C3} - \Delta\varphi_{C4}) & \sin(\Delta\varphi_{C1} - \Delta\varphi_{C2} + \Delta\varphi_{C3} - \Delta\varphi_{C4}) \\ 0 & 0 & -\sin(\Delta\varphi_{C1} - \Delta\varphi_{C2} + \Delta\varphi_{C3} - \Delta\varphi_{C4}) & \cos(\Delta\varphi_{C1} - \Delta\varphi_{C2} + \Delta\varphi_{C3} - \Delta\varphi_{C4}) \end{bmatrix}.$$

Note that the phase difference terms here are same as that of Eq. 3.

A single stage of Fig. 2 can be represented by *P2\*M\*P1\*I*, where I denotes the Stokes vector of the incident unpolarized light, that is [1, 0, 0, 0], and * denotes the matrix multiplication with P1=P(45°) and P2=P(-45°). Two stages can be represented in the same notation as *P1\*M\*P2\*M\*P1\*I*. It is to be noted that polarizer orientation is exchanged between P1 and P2 alternatively at every sequential stage. A four stage ASAP coronagraph can be represented by *P1\*M\*P2\*M\*P1\*M\*P2\*M\*P1\*I*. The output intensity, as plotted in Fig. 4, is given by the first element of this 4x1 matrix.

**References:**
1. Traub and Oppenheimer, "Direct Imaging of Exoplanets," Exoplanets, University of Arizona Press, 111-156, (2010).
2. Baudoz, Rabbia and Gay, "Achromatic Interfero Coronagraphy," Astron. & Astrophys. Suppl. Ser. 141, 319-329, (2000).
3. Barnaby R M Norris et al., "First on-sky demonstration of an integrated-photonic nulling interferometer: the GLINT instrument," MNRAS, 491, 4180–4193, (2019).
4. Murakami and Baba, "Common-path lateral-shearing nulling interferometry with a Savart plate for exoplanet detection," Optics Letters, 35, 18, (2010).
5. Kasdin et al., "Extrasolar Planet Finding via Optimal Apodized-Pupil and Shaped-Pupil Coronagraphs," Astrophys. J., 582, 1147, (2003).
6. Guyon et al., "Theoretical Limits on Extrasolar Terrestrial Planet Detection with Coronagraphs," Astrophys. J. Supp., 167, 81, (2006).
7. Kuchner and Traub, "A Coronagraph with a Band-Limited Mask for finding Terrestrial Planets," Astrophys. J., 570:900-908, (2002).
8. Rouan D. et al., "The Four-Quadrant Phase-Mask Coronagraph. I. Principle," PASP, 112:1479-1486, (2000).
9. Palacios, D. M., "An optical vortex coronagraph," Proc. SPIE, 5905, 196, (2005).
10. Webster Cash, "Detection of Earth-like planets around nearby stars using a petal-shaped occulter," Nature Letters, 442, 51–53, (2006).
11. Guyon et al., "Theoretical Limits on Extrasolar Terrestrial Planet Detection with Coronagraphs," Astrophys. J. Supp., 167, 81, (2006).
12. Mawet et al., "Review of small-angle coronagraphic techniques in the wake of ground-based second-generation adaptive optics systems." Proc. SPIE 8442, Space Telescopes and Instrumentation 2012: Optical, Infrared, and Millimeter Wave, 844204, (2012).
13. Hecht E. "Optics," Chap. 8, Addison-Wesley, (2002).
14. Francisco E. Veiras, Liliana I. Perez, and María T. Garea, "Phase shift formulas in uniaxial media: an application to waveplates," Appl. Opt. 49, 2769-2777, (2010).
15. Bracewell R N, "Detecting nonsolar planets by spinning infrared interferometer," Nature 274, 780–781, (1978).
16. Michael Bottom et al., "Stellar Double Coronagraph: A Multistage Coronagraphic Platform at Palomar Observatory," PASP 128 075003, (2016).
17. Pluznik et al., "Exoplanet Imaging with a Phase-induced Amplitude Apodization Coronagraph III. Hybrid Approach: Optical Design and Diffraction Analysis," Astrophys. J., 644, 1246, (2006).
18. Pueyo et al., "Design of Phase Induced Amplitude Apodization Coronagraphs over Square Apertures," Astrophys. J. Supp., 195, 25, (2011).
19. Ramachandran, G.N., "Birefringence of crystals and its temperature-variation," Proc. Indian Acad. Sci., 26, 77, (1947).
20. P. A. Miles-Páez et al., "Simultaneous optical and near-infrared linear spectropolarimetry of the earthshine," Astron. & Astrophy, 562, L5, (2014).
21. Jose Carlos del Toro Iniesta, "Introduction to Spectropolarimetry," Cambridge University Press, (2009).

***Biography:*** *Bhavesh Jaiswal is a scientist at Space Astronomy Group, U. R. Rao Satellite Centre (ISRO), India. He graduated from the Indian Institute of Space Science and Technology (IIST), India in 2011 with a degree in Physical Sciences and is currently working towards his PhD at the Indian Institute of Science (IISc), India. His research interest includes developing instrument concept for studying planets.*

Table 1: Various parameters considered for the 4 birefringent crystals C1, C2, C3 and C4 in the design of coronagraph in Fig. 2. Refer Fig. 1 and 2 for relevant geometry parameters.

|  | C1 | C2 | C3 | C4 |
|---|---|---|---|---|
| L | 50 mm | 50 mm | 50 mm | 50 mm |
| α | 45° | 45° | 10° | 10° |
| ϴ | 0° | 0° | 18° | 18° |
| δ | 90° | 0° | 90° | 0° |
| ϵ | 10″ | 10″ | 10″ | 10″ |
| Material | Calcite | Calcite | Calcite | Calcite |

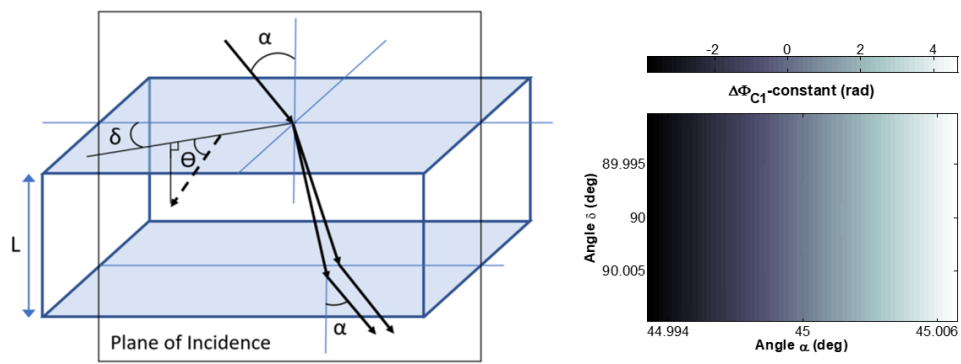

Figure 1: [Left] Ray transmission geometry in a thick birefringent crystal. The light ray, incident at an angle α, is marked as continuous arrows and the optic axis direction is marked as a dashed arrow. The rays are seen to split into 'ordinary' and 'extra-ordinary' rays at the exit of the crystal. [Right] The phase difference between extra-ordinary and ordinary ray given for a small range of angles α and δ. The configuration considered here is of crystal C1 used in Fig. 2.

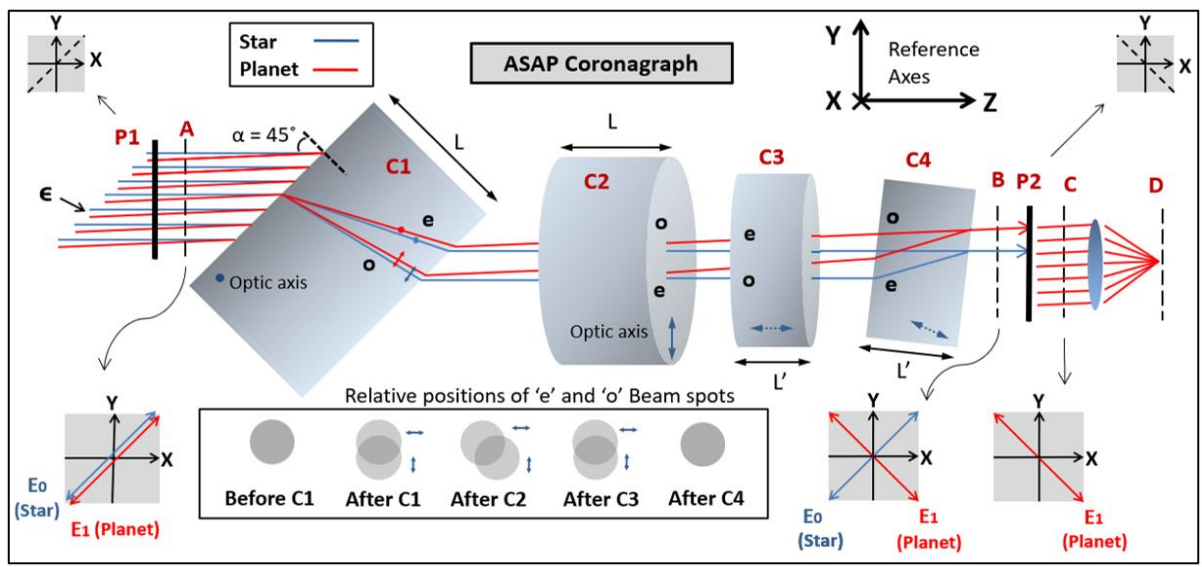

Figure 2: Functional diagram of the ASAP Coronagraph concept. Two light beams: star (blue) and planet (red), with an angle ϵ between them, enter from left (collimated beams after telescope secondary mirror) and pass through a set of parallel plate birefringent crystals C1, C2, C3 & C4. The reference axes are marked in the top of the figure. The crystals C1 and C2 are rotated by 45° about +X and -Y axes respectively. The crystals C4 and C3 are rotated by 10° about +X and -Y axes (not shown) respectively. The direction cosines of the optic axes of C1 to C4 are [1, 0, 0], [0, 1, 0],

[cos(108°+10°), 0, cos(18°-10°)] & [0, cos(108°+10°), cos(18°-10°)] respectively and is marked for each crystal. P1 and P2 are linear polarizers with their directions marked with dashed line in the X-Y plane at the top of the figure. Imaginary plane A, B & C are located at key positions to show the change in the state of polarization of star and planet light where $E_0$ and $E_1$ show the plane of polarization of star and planet light respectively at planes A, B and C in the bottom of the figure. After plane B the polarizer P2 blocks star light and lets the planet light pass through. Due to difference in the refractive index of 'ordinary' and 'extra-ordinary' beam inside the crystals the 'o' ray and 'e' refract at different angles which causes the beam spot of 'o' and 'e' ray to separate after passing through the crystal. The 'e' and 'o' rays are marked for all 4 crystals along with their plane of polarization. This separation of the two beam spots is shown in the bottom box. To get the desired phase difference in the entire beam area, it is required to overlap these 2 spots, and which is why the crystals C3 and C4 move both the spots in X and Y directions so that a perfect overlap is achieved after C4.

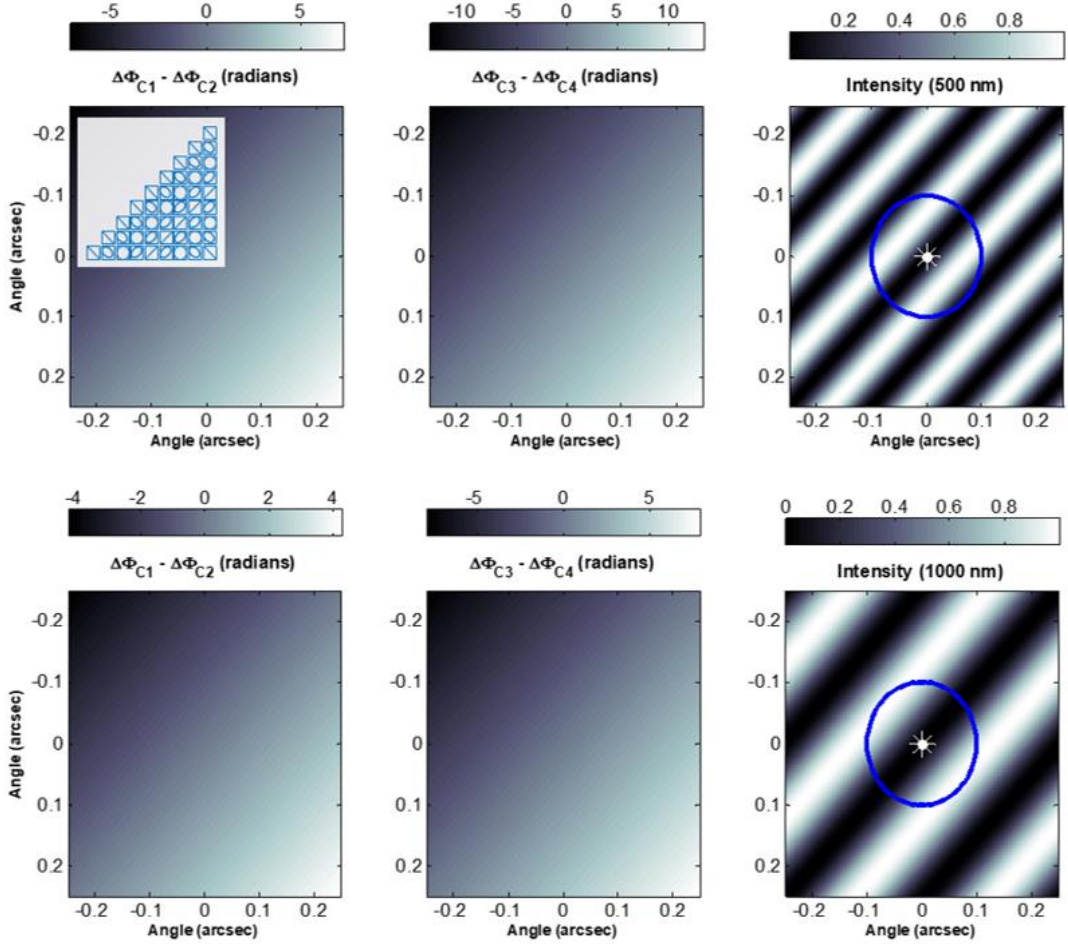

Figure 3: On-sky performance of the stellar coronagraph for a star-planet system located at a distance of 10 parsecs. The phase difference (as in Eq. 1) is shown on left at plane C of Fig. 2. The phase difference is plotted in units of 'radians'. After plane B and Polarizer P2 the angular sensitive performance is seen as fringes on the sky plane in right-panel. The simulations are shown for two wavelengths 500 and 1000 nm. The location of star and a 1 au orbit is shown in the right panel. The star is located at the dark-fringe irrespective of the wavelength. The inset figure in Phase difference plot shows how the state of polarization changes with phase, from linear to elliptical to circular and then back to elliptical to linear.

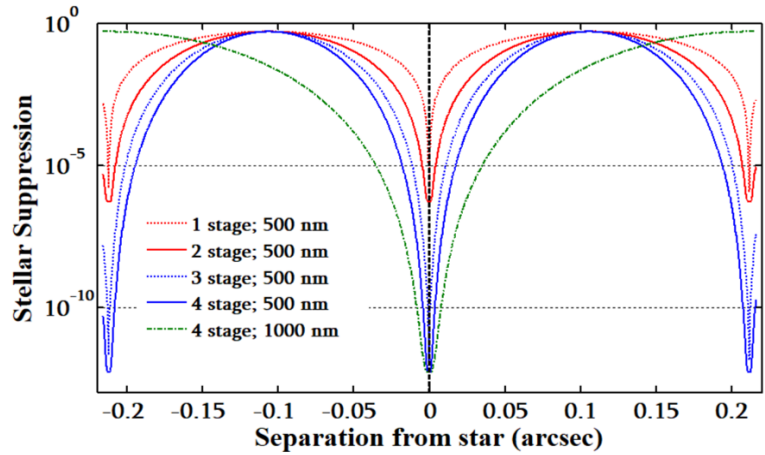

**Figure 4:** Star light suppression or contrast achieved in ASAP coronagraph for single and multiple stages. The simulations consider the realistic polarizers with an extinction ratio of $10^6:1$. The stellar diameter is marked by a vertical line at the centre whose thickness is equivalent to stellar diameter for a Sun like star at 10 parsecs.

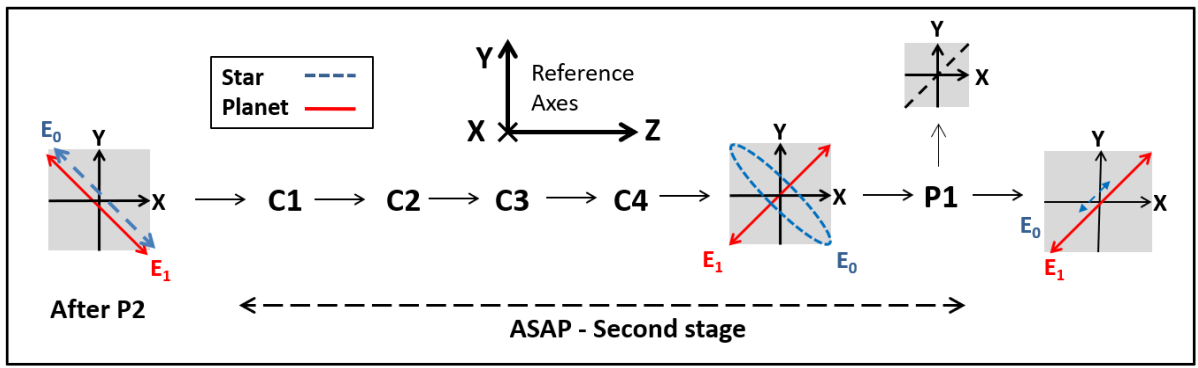

**Figure 5:** Representation of the second stage (after P2 of first stage) to reduce the residual star light from first-stage. This is same as the configuration of first stage as shown in Fig. 2 except for the polarizer P1 instead of P2. The amplitude of residual star light ($E_0$) is seen to decrease after P1.

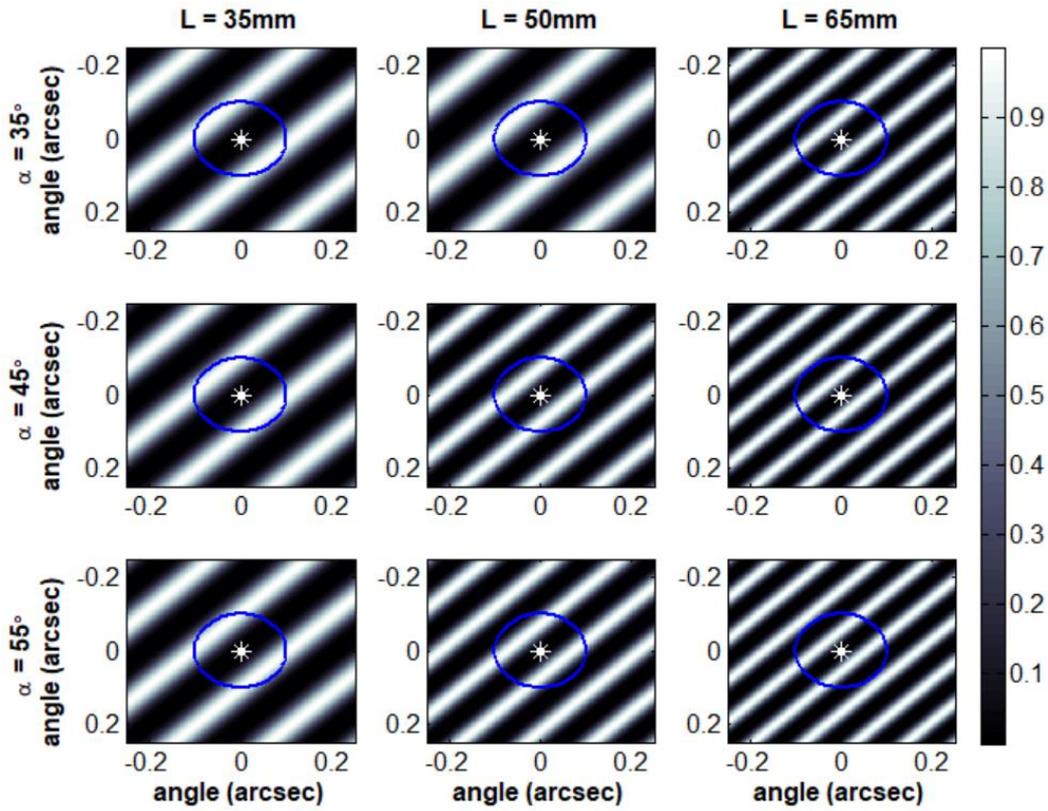

**Figure 6:** Sensitivity of the Intensity fringe spacing (as in Fig. 3) to thickness of crystals 'L' and the incident angle 'α'. Similar to Fig. 3 the orbit of Earth at 1 au at 10 parsecs is marked as a circle.

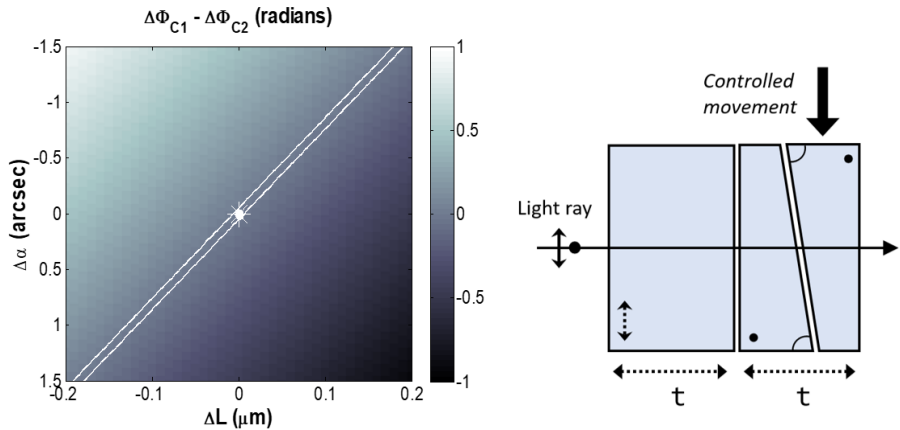

**Figure 7:** [Left] Sensitivity of the phase differences to small relative differences in the incident angle 'α' and crystal thickness 'L' of the two crystal pairs. Phase differences (given in radians) due to C1 and C2 are shown on left. Two parallel white lines show the desired tolerances (centred at zero phase difference) with respect to stellar diameter. [Right] A configuration of Babinet-Soleil compensator used to compensate for extra path length by the sliding one of the crystals. The optic axis directions are marked inside the crystals.